\documentclass[11pt, a4paper]{article}
\pdfoutput=1
\usepackage{jheppub}
\usepackage{graphicx}
\usepackage{epsfig,amsmath,amsthm}
\usepackage{epstopdf}

\newcommand{\be}{\begin{equation}}
\newcommand{\ee}{\end{equation}}
\newcommand{\bea}{\begin{eqnarray}}
\newcommand{\eea}{\end{eqnarray}}
\def\bse{\begin{subequations}}
\def\ese{\end{subequations}}
\newcommand{\IR}{\mathbb{R}} 
 
\def\IZ{\relax\ifmmode\hbox{Z\kern-.4em Z}\else{Z\kern-.4em Z}\fi}

\newcommand{\non}{\nonumber \\}

\def\del{{\partial}}

\def\bi{\begin{itemize}} \def\ei{\end{itemize}}

\def\({\left(} \def\){\right)}
\def\[{\left[} \def\]{\right]}
\def\<{\left<} \def\>{\right>}

\title{The algebraic locus of Feynman integrals}

\author{Barak Kol  \\
{\it Racah Institute of Physics, Hebrew University, Jerusalem 91904, Israel} \\
{\tt barak.kol@mail.huji.ac.il}
}

\abstract{In the Symmetries of Feynman Integrals (SFI) approach, a diagram's parameter space is foliated by orbits of a Lie group associated with the diagram.  SFI is related to the important methods of Integrations By Parts and of Differential Equations. It is shown that sometimes there exist a locus in parameter space where the set of SFI differential equations degenerates into an algebraic equation, thereby enabling a solution in terms of integrals associated with degenerations of the diagram. This is demonstrated by the two-loop vacuum diagram, where the algebraic locus can be interpreted as a threshold phenomenon and is given by the zeros of the Baikov polynomial. 
 In addition, the symmetry group is interpreted geometrically as linear transformations of loop currents which preserve the subspace of squared currents within the space of quadratics in loop currents.}

\begin{document}
\maketitle

\hfill \begin{tabular}{l}
To freedom \\ 
at passover \\ 
the holiday of freedom. 
\end{tabular}

\vspace{.5cm}

\section{Introduction}

Integration By Parts (IBP) is an important method for computing Feynman integrals \cite{ChetyrkinTkachov81}, namely scalar integrals associated with Feynman diagrams. In some cases it allows to reduce the computation of a Feynman integral to the computation of simpler integrals, known as ``master integrals''. An algorithm \cite{Laporta2001} was devised to implement this method, and  several associated computer packages exist, see e.g. \cite{Smirnov2013dia} and references therein.  

The method is widely used and currently the original paper \cite{ChetyrkinTkachov81} has collected over 1,000 INSPIRE citations. However, it still lacks an ``instruction manual'', namely, given a diagram it is not known how much simplification to expect of the method (prior to applying the computer package). 

Recently, a geometry underlying the method was exposed \cite{SFI}. This approach considers the dependence of the integrals on its parameters,  namely masses and scalars composed of external momenta. It turns out that in parameter space the Integration By Parts method produces a set of differential equations (related to the Differential Equations method, see the review \cite{HennRev2014} and references therein). More precisely, this set is made of certain ``numerator free'' IBP relations, together with additional IBP-like relations which should be added in the presence of external legs. The set of differential equations defines a continuous (Lie) group $G$ acting on parameter space, which is thereby foliated into $G$-orbits. Within each $G$-orbit the dependence of the integral on its parameters was shown to reduce to a certain line integral. The integrand for the latter is expressed in terms of integrals associated with degenerated diagrams/integrals.

In this paper, we show how sometimes there exists a variety in parameter space where the set of differential equations reduces to an algebraic equation. We call this variety the ``algebraic locus'' . On it the integral under study can be expressed in terms of degenerate integrals.

We start in section \ref{sec:SFI} by reviewing the basics of the SFI approach, and offering a geometrical interpretation of the associated group. In section \ref{sec:locus} we give a general description of the algebraic locus phenomenon and provide a general method to present its location. These ideas are demonstrated in section \ref{sec:demonstrate} by the two-loop vacuum diagram. Finally section \ref{sec:summary} is a summary and a discussion.

\section{Symmetries of Feynman Integral (SFI) approach}
\label{sec:SFI}

Here we briefly describe the set-up, culminating with the equation set from \cite{SFI} and its novel geometrical interpretation to be described and  demonstrated in \cite{VacuumSeagull}.

Given a Feynman diagram $\Gamma$ we define the associated Feynman integral by \be
I_\Gamma(x) := \int \frac{dl}{\prod_i \(k_i^2-x_i\)}  
 \label{def:Ix}
\ee
where $dl := \prod_{r=1}^L d^dl_r$ denotes integration over all loops for a general space-time dimension $d$; the index $i=1,\dots,P$ runs over all propagators; $k_i^\mu$ is the energy-momentum of the propagator and it should be considered to be expressed as a combination of loop and external momenta;  for $i=1,\dots,P$ the $x_i$ parameters can be identified with squared masses $x_i \equiv m_i^2$; and finally for $i=P+1,\dots$ the $x_i$ parameters are (independent) scalars composed of the external momenta. The parameters $x_i ~ i=1,\dots,P$ have an alternative interpretation as formal parameters of a partition function associated with propagator indices (powers), see \cite{SFI}. Note that  \emph{they are conjugate to the Schwinger parameters under a Laplace transform} (this is implied by the definitions). 

{\bf Geometric interpretation: freedom of loop variables} \cite{VacuumSeagull}. As usual there is some freedom for choosing the loop variables. The essential idea can be stated for vacuum diagrams. One makes a choice of loop variables and considers a generator of a linear re-definition of loop variables \be
\delta l_r = T_r^{~s}\, l_s ~, \qquad r,s=1,\dots,L 
\label{glLaction}
\ee
where $T_r^{~s}$ is a matrix, namely $T \in gl(L)$ the algebra of the general linear group.\footnote{
Such an infinitesimal transformation is not part of the usual freedom of choosing loop variables, but it can be considered a generalization thereof, allowing not only for integer coefficients in the re-definition, but also for real ones.}
 Upon acting on the integral (\ref{def:Ix}) a sub-algebra of generators $g \subset gl(L)$ would produce partial differential equations for $I(x)$. In \cite{SFI} the elements of $g$ were termed ``numerator free'' IBP relations. It can be described geometrically as follows. Consider the space spanned by all squared currents, namely 
\be
S := Sp \left\{ k_i^2 \right\}, \qquad i=1,\dots,P 
\label{def:squares}
\ee
Since each propagator current $k_i$ is a linear combination of loop currents $l_r$ we have 
\be
S \subset Q
\label{SinQ}
\ee
 where $Q$ are all the quadratic expressions in the loop currents \be
 Q:=S^2(\mbox{loop currents}) \equiv Sp \left\{ l_r \cdot l_s \right\} \qquad r,s=1,\dots,L
 \label{def:Q}
 \ee
The action of $GL(L)$ on the loop variables induces an action on $Q$. \emph{The numerator free group $G$ is precisely the $GL(L)$ subgroup which preserves $S$ within $Q$.} This follows straightforwardly from the definitions. The geometric interpretation of $G$ will be further discussed and demonstrated in \cite{VacuumSeagull}.

In this way the mapping $\Gamma \to G$ which maps any Feynman graph $\Gamma$ to a Lie group $G$ is factored through the pair $S \in Q$, namely \be
\Gamma ~~\to~~ S \subset Q ~~\to~~ G
\label{factoring}
\ee
In particular if two different graphs, generate $(S,Q)$ pairs which are isomorphic under $GL(L)$, then they will have the same G.

{\bf Equation set}. The resulting relations form a set of linear first-order partial differential equations  \cite{SFI} \bea
	0 &=&  D^A I + J^A  \non
	D^A &:=&  c^A + \(T^A\)^i_{~j}\, x_i \, \del^j ~.
\label{IBPxa}
\eea
The equations are labelled by $A=0,1,\dots$ and $J^A$ are source terms expressed by integrals associated with degenerations of $\Gamma$. The unit generator $l_r\, \del^r$ is always numerator free, namely it belongs to $G$. The associated equation is  \be
0 = \[ c_0-  x_i \, \del^i \] I 
\label{dim-eq}
\ee
 where $c_0$ is the $x$-dimension of $I$.  It can be interpreted as the Euler identity for homogeneous functions, namely as stating the $x$-dimension of $I$ (dimensional analysis). We choose to number it as the zeroth equation $A=0$. The remaining generators $a=1,2,\dots$ are chosen such that $c^a=0$ and thus generate a normal subgroup of $G$ which was termed the constant free subgroup and was denoted by $G_2$ (not to be confused with the simple group with the same name), namely \bea
 	\(T^0\)^i_{~j} &=& - \delta^i_{~j} \qquad J^0 = 0 \non
 	c^a &=& 0 ~~ a=1,2,\dots
	\label{G2-relations}
 \eea

\begin{figure}
\centering \noindent
\includegraphics[width=8cm]{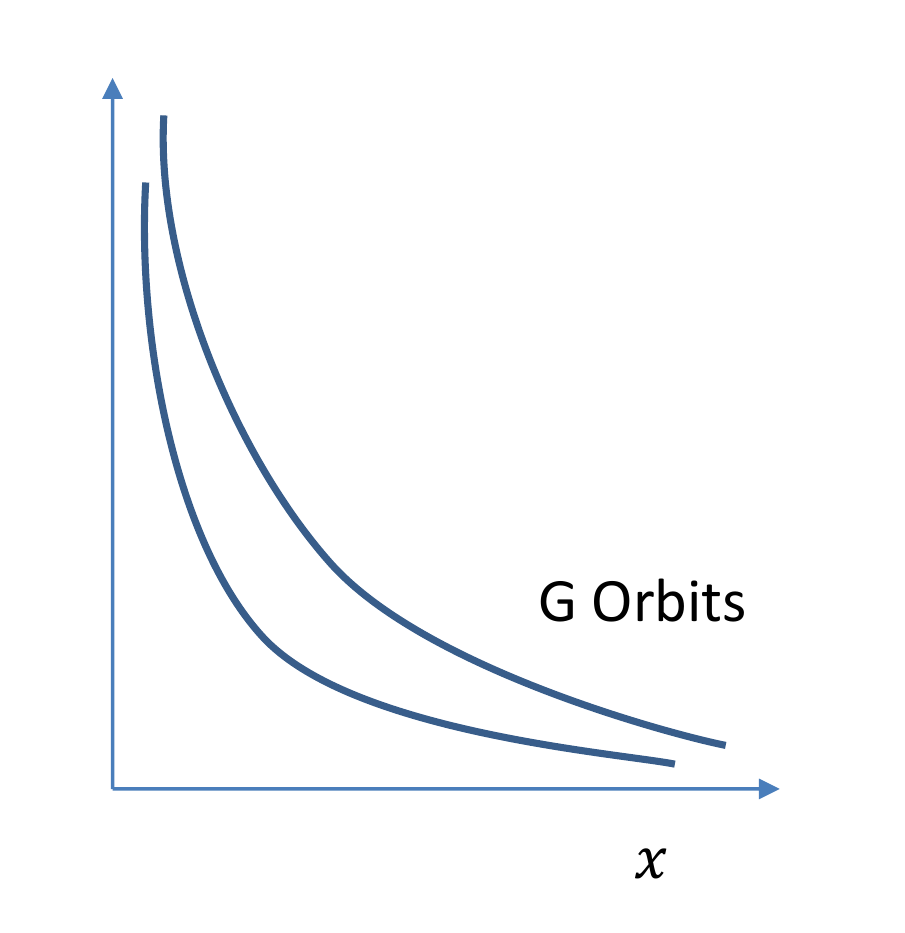}
\caption[]{This figure demonstrates the main result of the Symmetries of Feynman Integrals (SFI) approach. A Feynman diagram defines a group $G$ which acts on the parameter space of the diagram which is denoted by $x$ variables. Accordingly, the $x$ space is foliated into $G$ orbits. The method allows to reduce the dependence of $I$ within an orbit to a line integral.}
 \label{fig:central}
\end{figure}

The main result of the SFI approach is to foliate the the parameter space into $G$ orbits, such that the dependence of $I$ within an orbit can be reduced to a line integral. This is represented schematically in fig. \ref{fig:central}. The line integral is over edge-contraction degenerations of the original diagram. These diagrams, in turn, may also have a representation through a line integral thereby turning the original representation into an iterated diagram. \emph{I wonder whether this could be the origin of the common appearance in evaluations of Feynman Integrals of Multiple PolyLogarithms and other special functions involving iterated integrals.}

\section{Algebraic locus}
\label{sec:locus}

As usual, the matrices $T^A$ are chosen to be linearly independent, namely \be
 \alpha_A\, T^A = 0 \implies \alpha_A=0 ~ \forall A
 \ee
 
However, at any specific point in $x$ space, the vectors $\(v^A\)_j = \(T^A\)^i_{~j}\, x_i$ belong to the tangent space to parameter space and may very well be linearly dependent, namely for some points $x$ there exist $\alpha_A(x)$ not all of them vanishing, such that \be
\alpha_A(x)\, \(T^A\)^i_{~j}\, x_i = 0 ~.
\label{pointwise-relation}
\ee
 We refer to such relations as ``pointwise linear dependence''.

According to (\ref{G2-relations})  $\alpha_A(x)\, c^A = \alpha_0(x)\, c_0$ and hence the linear combination of the equation set (\ref{IBPxa}) with coefficients $\alpha_A(x)$ at $x$ yields \be
0=\alpha_0(x)\, c_0 \, I + \alpha_a(x) J^a
 \label{alg-eq}
\ee  
Here the differential equations canceled out and an algebraic relation remains. As long as \be 
\alpha_0(x) \neq  0
 \label{alpha0_neq_0}
\ee
and recalling that for generic spacetime dimension $d$ $c_0 \neq 0$ this equation can be immediately solved for $I$ in terms of the sources $J^a$ namely \be
	I_{\rm alg. locus}  = - \frac{ \alpha_a(x)\, J^a}{\alpha_0(x)\, c_0} ~.
	\label{alg_loc_soln}
	\ee
Hence we refer to the locus in $x$ space where (\ref{pointwise-relation},\ref{alpha0_neq_0}) holds as the \emph{algebraic locus} of the equation set. This is one of the main results of this paper.

Note that the condition for pointwise linear dependence (\ref{pointwise-relation}) can be further reduced through splitting the summation into $A=0,a$ and using (\ref{G2-relations}) to obtain \be
  \alpha_a(x)\, \(T^a\)^i_{~j}\, x_i = \alpha_0(x) x_j 
  \ee
  This can be interpreted as a local degeneration of the $G_2$ generator $T:=\alpha_a(x)\, T^a$ into the dimension generator $x=x_i \del^i$. Furthermore, this is an eigenvector condition for $T$. Here we are mostly interested in real eigenvectors $x_i$ and real eigenvalues $\alpha_0$. It might be interesting to consider the effect of non-real eigenvalues on the analytic continuation of $I(x)$. 
 
 We would like to comment on another kind of pointwise linear relations, those where $\alpha_0=0$ in (\ref{pointwise-relation}). In this case (\ref{alg-eq}) cannot be solved for $I(x)$ and it is not part of the algebraic locus. Rather, it becomes an algebraic constraint among the sources $J^a$. Such a constraint should be distinguished from the integrability constraints on the sources, eq. (4.6) in \cite{SFI} which differ by being differential. 

\subsection*{Equations describing the algebraic locus} 

The algebraic locus was defined in the previous subsection as the locus in $x$ space where the radial direction $x= x_i \del^i$ is within the subspace of the tangent plane spanned by $G_2$. This locus can be described by a set of equations as follows.

One starts by seeking a homogeneous solution $I_0$ to the equation set (\ref{IBPxa}). By definition, $I_0$ is constant over orbits of $G_2$, hence it can depend only on $G_2$ invariants $P_1,P_2,\dots$. In order to obtain expressions for the invariants one starts by choosing a maximal number $M$ of pointwise independent $G_2$ vector fields $v_a = \(T^a\) ^i_{~j} x_i \del^j$ and taking their exterior product $\bigwedge_{a=1}^M v_A$ to achieve a multi-vector describing the tangent plane to the $G_2$ orbit ($M$ is the orbit's generic dimension). Next one defines its dual 
form \be
\omega := * \( \bigwedge_{a=1}^M v_a \)
\label{def:om}
\ee
Since the family of $M$-planes is integrable to a $G_2$ orbit I believe the surface could be written (at least locally) as a complete intersection of $G_2$ invariants and hence $\omega$ should have the form \be
 \omega = f \bigwedge_r dP_r
 \label{om-form}
 \ee
where $f$ is some function which we shall determine. Note that while $\omega$ depends on the choice of the $M$ generators, the invariants are independent of this choice and so the dependence is absorbed by  $f$. To determine $f$ one computes the differential $d\omega$ which according to  (\ref{om-form})  should satisfy \be
d\omega = d\log f \wedge \omega 
\label{domega}
\ee
The last equation is an over-constrained equation set for $d\log f$ whereby $f$ can be obtained. 

Summarizing, the $G_2$ invariants $P_1,\, P_2, \dots$ can be obtained through \be
\bigwedge_r dP_r = * \( \bigwedge_{a=1}^M v_a \) /f
\label{invar}
\ee
where $f$ is solved through (\ref{domega}). This is one of the main results of this paper. A demonstration of this process will be given in the next section.

Once the invariants are known we can proceed to determining equations for the algebraic locus \be
0 = x \cdot \bigwedge_r dP_r  
\label{eq:alg-loc}
\ee
In the presence of a single invariant it becomes \be
0 = x \cdot dP_1  = \Delta_1\, P_1
\ee
where the Euler identity for homogenous functions was used  and $\Delta_1$ is the $x$ dimension of $P_1$. So that the algebraic locus is given by the equation \be
P_1=0
\label{codim1locus}
\ee
In the presence of two invariants  (\ref{eq:alg-loc}) becomes \be
 0 =  x^i\, \( \del_i P_1\, \del_j P_2 -  \del_i P_2\, \del_j P_1 \) = \Delta_1\, P_1\, \del_j P_2 - \Delta_2\, P_2\, \del_j P_1
\ee
Hence the algebraic locus is obtained when either $P_1=P_2=0$ or $P_1=\del_j P_1=0$ or a similar condition for $P_2$.  This list of conditions is equivalent to requiring a double zero of $P_1\, P_2$.

\section{Demonstration: the two-loop vacuum diagram}
\label{sec:demonstrate}

\begin{figure}
\centering \noindent
\includegraphics[width=15cm]{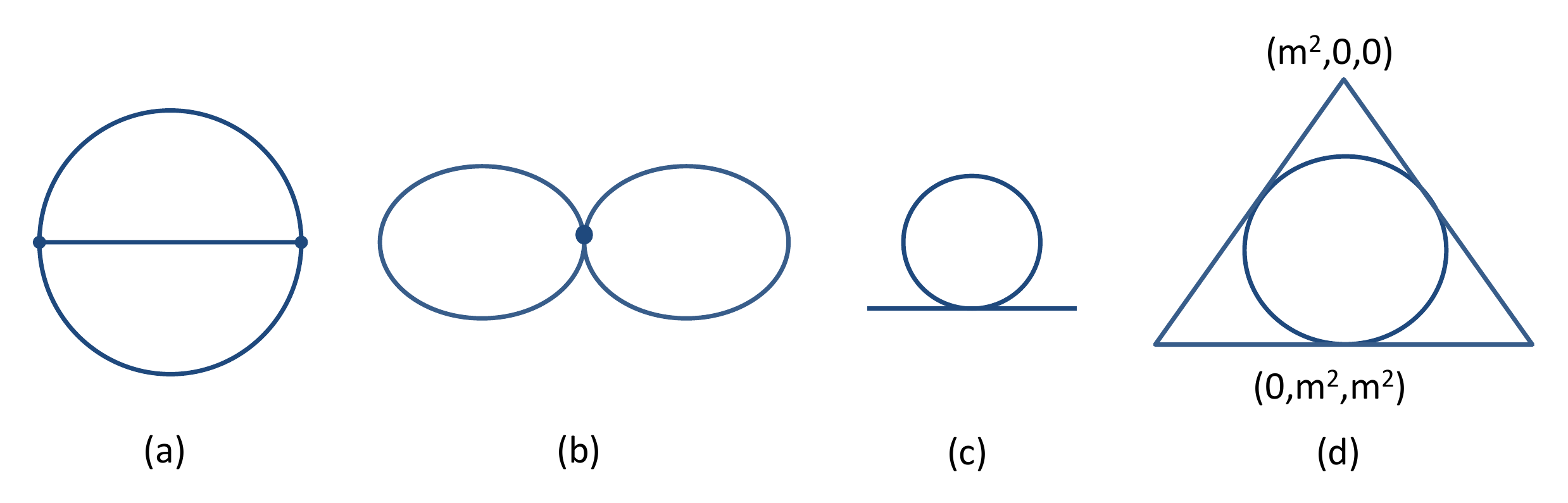}
\caption[]{(a) The 2-loop vacuum diagram, or the diameter diagram. (b) The unique degeneration of the diameter diagram.  (c) A factor of the previous diagram which defines the source function $j_i$ (\ref{def:ji}). (d) The parameter space for diag. (a) can be illustrated as 2-simplex, namely an equilateral triangle, after factoring out for scale. At each vertex a single propagator is massive, on the edges a single propagator is massless, and at the edge midpoints the other two have equal masses. The inscribed circle is the algebraic locus.}
 \label{fig:vacuum}
\end{figure}

In this section we shall study the algebraic locus of the integral associated with the two-loop vacuum diagram shown in  fig. \ref{fig:vacuum}(a), which can also be called the diameter diagram. It is the same case which was used to demonstrate the SFI approach in \cite{SFI}. 

This diagram consists of 3 propagators hence the integral $I(x)$ depends on 3 mass-squares $x_1,\, x_2,\, x_3$. For a specific choice of loop currents its definition becomes \be
I\(x_1,\, x_2,\, x_3\) = \int \frac{d^dl_1\, d^dl_2}{\( l_1^2-x_1\) \( l_2^2-x_2\) \( (l_1+l_2)^2-x_3\)}
\label{def:Ix-Ln20}
\ee

The IBP equation set for this diagram consists of 4 equations which were brought in \cite{SFI} to a form symmetric with respect to the diagram's discrete symmetry group, an $S_3$ permuting the propagators  \be
 0 = \[ \begin{array}{cccc}
 d-3 	& -x_1		& -x_2		& -x_3 	\\
 0 		& x_2-x_3	& x_2		& -x_3 	\\
 0 		& -x_1 		& x_3-x_1	& x_3	\\
 0		& x_1 		&-x_2		& x_1- x_2 \\
 \end{array} \] \,
 \[ \begin{array}{c}
1 \\
\del_1 \\
\del_2 \\
\del_3 \\
\end{array} \] \, I +a
\[ \begin{array}{c}
0 \\
(j_3-j_2)\, j_1' \\
(j_1-j_3)\, j_2' \\
(j_2-j_1)\, j_3' \\
\end{array} \]
\label{def:set-Ln20}
\ee
Here the functions $j_i$ which appear as sources are given by the Feynman integral associated with the diagram in fig. \ref{fig:vacuum}(c) \be
j_i :=j (x_i) \qquad j(x) = I_{fig. \ref{fig:vacuum}(c)} 
\label{def:ji}
\ee
The function $j(x)$ originates from the sole degeneration of the diameter diagram shown in fig. \ref{fig:vacuum}(b) which in turn factorizes into two fig. \ref{fig:vacuum}(c) diagrams. Since $j(x)$ is given by a power law we have $x\, j'(x) = \frac{d-2}{2}\, j(x)$ .

The last three equations in (\ref{def:set-Ln20}) define a representation of the generators of the constant-free sub-algebra $G_2$, which in this case is $SL(2,\IR)$ in its 3 dimensional irreducible representation.

{\bf Algebraic locus}. The $G_2$ invariant was already mentioned in \cite{SFI} eq. (6.15), yet here we shall derive it systematically using the method from the previous section. From (\ref{def:set-Ln20}) we define the vector fields \bea
v_1 &:=& \( x_2 - x_3 \) \del_1 + x_2\, \del_2 - x_3\, \del_3 \\
v_2 &:=& -x_1\, \del_1 + \( x_3 - x_1\) \del_2 + x_3\, \del_3 \\
v_3 &:=& x_1\, \del_1 - x_2 \, \del_2 + \(x_1 - x_2 \) \del_3
\label{def:v}
\eea
One confirms that the maximal number of non-vanishing wedge product is 2, namely the dimension of the $G_2$ orbit is 2. Let us choose $v_1$ and $v_2$ and we find the dual form $\omega$ to be the 1-form  \be
\omega = * \( v_1 \wedge v_2 \) = x_3 \left[ \( x_2 + x_3-x_1 \) dx_1 + \( x_1 + x_3 - x_2 \) dx_2 + \( x_1 + x_2 - x_3\) dx_3 \right]
\ee
By computing $d\omega$ as in (\ref{domega}) we can identify that $f=x_3$ and finally the invariant is \be
 P \equiv \lambda =  2 \(x_1^{~2}  + x_2^{~2}  +x_3^{~2} \) - \( x_1 + x_2 + x_3 \)^2 
 \label{def:lambda}
\ee
  where $\lambda\(m_1^{~2}, m_2^{~2} , m_3^{~2}\)$ is the K\"all\'en function which appears in relativistic 3-body kinematics \cite{WikiKallen}. More precisely, in the decay process $m_1 \to m_2 + m_3$ one has $\lambda = 4\, m_1^{~3}\, P^2$ where $P^2$ is the momentum squared of particle 2 (and of particle 3) in the center of mass frame. Interestingly, $\lambda$ is symmetric under permutations of all three particles.
 
Note that the $\lambda$ invariant is a quadratic in the $x$'s and has a $(2,1)$ signature (positive,negative). Accordingly the $x$ space can be identified with a $2+1$ Minkowski space-time and its symmetry group $G_2=SL(2,\IR)$ can be identified with the Lorentz transformations $SO(2,1) \simeq SL(2,\IR)/\IZ_2$.

Note also that the $\lambda$ invariant coincides with the Baikov polynomial of this diagram \cite{Baikov1996} known as ``the basic polynomial $P(x)$'' \cite{SmirnovBook2006}, which is obtained as a determinant of a certain $x$ dependent matrix (described explicitly in \cite{SFI}).

By the general argument for a codimension 1 $G_2$ orbits (\ref{codim1locus}), the algebraic locus is given by \be
0= \lambda \equiv  2 \(x_1^{~2}  + x_2^{~2}  +x_3^{~2} \) - \( x_1 + x_2 + x_3 \)^2
\label{locus2L}
\ee 
 The parameter space and the algebraic locus within it are shown in fig. \ref{fig:vacuum}(d). The parameter space $(x_1,\, x_2,\, x_3)$ is 3 dimensional, yet dimensional analysis makes the radial dependence of $I$ straightforward. Therefore the essential dependence is on the the projective space, which can be realized concretely as the $x_1+x_2 + x_3=1$ plane. We assume all $x_i$ to be positive being mass squares (in fact for negative $x$'s the source terms in the SFI equations become problematic).  In the projective space the boundary of the wedge  $x_1,\, x_2,\, x_3 \ge 0$ (where one of the propagators in massless) becomes the triangle (or 2 -simplex) shown in the figure. The algebraic locus (\ref{locus2L}) is the lightcone within the Minkowski space interpretation and it becomes a circle after projection. The circle is inscribed inside the triangle, being tangent to it  it at points where two of the $x$'s (or masses) are equal and the third is massless. 

{\bf Solution at locus}. By construction, on the algebraic locus it is possible to take a linear combinations of the equations in  (\ref{def:set-Ln20}), such that an algebraic equation remains. Here it can be achieved by multiplying the equation set on the left by the row vector \be
\[ \begin{array}{cccc}
\alpha_0, & 	\alpha_1, & 	\alpha_2, & 	\alpha_3
\end{array} \] = 
 \[ \begin{array}{cccc}
x_1 + x_2 + x_3, & 	x_2 - x_3,  & x_3 - x_1, 	& x_1 - x_2 
 \end{array} \]
\label{alpha2L}
 \ee
In this way we obtain the solution for the Feynman integral \be
I \left|_{\lambda=0} \right. = \frac{1}{\( d-3 \) \( x_1+x_2 + x_3 \)} \[ \(x_2 - x_3 \) \(j_2 - j_3 \) j_1' + cyc. \]
\label{result}
\ee
 where $j_i$ was defined in (\ref{def:ji}) This is the main result of this section. \footnote{
 Anecdotally we note that it was recorded in the research notebook on 9 July 2015 at the Zurich airport during the return trip from Amplitudes 2015.}

Note that this form of the solution is not unique. Since the last 3 equations of (\ref{def:set-Ln20}) are point-wise linearly dependent (an algebraic constraint) one could add to the $\alpha_A$ row vector (\ref{alpha2L}) any multiple of \be
 \[ \begin{array}{cccc}
0,	 & 	x_1,  & x_2, 	& x_3 
 \end{array} \]
 \ee
The expression given in (\ref{result}) was chosen for its simple symmetry under the $S_3$ symmetry which permutes the variables $x_1, x_2, x_3$.

Note that while $\lambda\(x_1,x_2,x_3\)$ (\ref{def:lambda}) cannot be factorized when considered as a polynomial in the $x  \equiv m^2$ variables, it does admit a factorization in terms of the mass variables \be
 \lambda\(m_1^2, m_2^2, m_3^2\)  = \( m_1-m_2-m_3 \)  \( m_2-m_3-m_1 \)   \( m_3-m_1-m_2 \)   \( m_1 + m_2 + m_3 \)  
\label{threshold-factors}
\ee
Accordingly the algebraic locus circle (fig. \ref{fig:vacuum}d) is divided into three segments: $m_1=m_2+m_3$ and permutations thereof. The equation $m_1=m_2+m_3$ denotes a threshold phenomenon which is associated here with the algebraic locus. 

{\bf Tests}. The point $x_1=x_2=m^2, \, x_3=0$ is on the algebraic locus. In fact it is the point where it intersects the massless triangle. At this point (\ref{result}) becomes \be
I \( x_1=x_2=m^2, \, x_3=0 \) = \frac{d-2}{2(d-3)\, m^2}\,  j^2(m^2) 
\ee
This result agrees with \cite{SmirnovBook2006} (2006) p. 45. 

More generally it will be shown in \cite{ExactDiameter} that (\ref{result}) agrees over its whole domain of variables with the exact and general result for the diameter diagram \cite{DavydychevTausk1992}.

\section{Summary and Discussion}
\label{sec:summary}

In this paper we found that sometimes the SFI (Symmetries of Feynman Integrals \cite{SFI}) equation set has a special locus in parameter space where the equations turn algebraic rather than differential. This was termed the algebraic locus. Within it the integral of interest reduces to integrals associated with degenerations of the original diagram. From the group perspective this happens at points in parameter space where a vector field associated with a constant-free generator becomes radial (a direction associated with the dimension generator). Note that \emph{the algebraic locus is also the place where the standard IBP recursive relations imply the same reduction into degenerations.}

This phenomenon was demonstrated by the case of the two-loop vacuum diagram, or the diameter diagram, fig. \ref{fig:vacuum}(a). The algebraic locus was found to be the ``lightcone'' (\ref{locus2L}) drawn as a circle in  fig. \ref{fig:vacuum}(d). Within this locus the integral is given by the simple expression (\ref{result}).

The algebraic solution for the 2-loop vacuum diagram (\ref{result}) is certainly correct as it was tested against the general solution  \cite{DavydychevTausk1992}. It is novel in the sense that on this locus the general expression involving hypergeometric functions simplifies considerably, and neither the simplified form nor the associated locus appear to be present in the literature.  Interestingly this diagram together with the general analysis (\ref{alg_loc_soln},\ref{invar}) (and additional examples including  \cite{VacuumSeagull}) suggest that algebraic loci are a general phenomena for Feynman integrals.

We also mentioned a geometrical interpretation of the SFI group as the linear transformations over loop currents which preserve the space of squared currents within the space of quadratics in loop currents (\ref{factoring}).

\subsection*{Discussion.}

{\bf Built-in boundary conditions}. Usually differential equations must be supplemented by boundary / initial conditions. The SFI equation set was seen to imply the value of the solution at some locus. As such it suggests that at least part of the boundary conditions are built into this equation set, and need not be specified independently. 

{\bf Possible relations of the algebraic locus with other concepts}. A relation with threshold phenomena was observed around (\ref{threshold-factors}). A relation with the Baikov polynomial was observed above (\ref{locus2L}).

\subsection*{Acknowledgments}

I would like to thank Philipp Burda, Ruth Shir and Erez Urbach for collaboration on related projects and for comments on the manuscript; Ruth Lawrence-Naimark, Zlil Sela and Jake Solomon from the Institute of Mathematics at the Hebrew University for a discussion; and Johannes Henn and Vladimir Smirnov for correspondence.

This research is part of the Einstein Research Project ``Gravitation and High Energy Physics", funded by the Einstein Foundation Berlin, and it was also partly supported by the Israel Science Foundation grant no. 812/11 and the ``Quantum Universe'' I-CORE program of the Planning and
Budgeting Committee.  This line of research was mentioned in the ISF 681/15 research proposal which was refused for funding in March 2015.

\bibliographystyle{unsrt}

\end{document}